# Digitalizing Uncertain Information

Chris Partridge
*BORO Solutions Ltd*
*University of Westminster*
London, UK
0000-0003-2631-1627

Andrew Mitchell
*BORO Solutions Ltd*
*University of Westminster*
London, UK
0000-0001-9131-722X

Andreas Cola
*Telicent Ltd*
London, UK
0009-0008-8684-1498

*Abstract*—The paper sketches some initial results from an ongoing project to develop an ontology-based digital form for representing uncertain information. We frame this work as a journey from lower to higher levels of digital maturity across a technology divide. The paper first sets a baseline by describing the basic challenges any project dealing with digital uncertainty faces. It then describes how the project is facing them. It shows firstly how an extensional ontology (such as the BORO Foundational Ontology or the Information Exchange Standard) can be extended with a Lewisian counterpart approach to formalizing uncertainty that is adapted to computing. And then it shows how this is expressive enough to handle the challenges.

## I. Introduction

Uncertainty is a well-known and difficult problem in the wider intelligence community. In this short paper, we sketch some initial results from an ongoing project to formalize digital uncertainty information – or, more exactly, to design a sufficiently expressive ontology-based digital form (aka formal data infrastructure) to represent this uncertainty. We frame this work as a journey from lower to higher levels of digital maturity across a technology divide – one that requires building the ontology-based formal data infrastructure needed to support a migration from unstructured text-based to structured digital information. The motivation for this work is the expectation that having a clearer foundation for uncertain information will enable us to work more efficiently with it at scale.

When intelligence is couched in language, it often reflects an inherent uncertainty. The difference between saying that "it is possible that Anne was in Edinburgh" and "Anne was in Edinburgh", is, as the language indicates, that in the first case that we are uncertain in the second case that we are not. When intelligence is stored as data in an informational intelligence system, the system should be able to clearly respond to queries in a digital language that can reflect these kinds of underlying uncertainty – and certainty.

Achieving this digital formal clarity about uncertainty raises basic challenges which many current approaches find challenging. We outline the major ones in the paper. We characterize each challenge. We use the same simple 'use case' to illustrate them: "it is possible that Anne was actually in Edinburgh last Saturday, 30th March 2024". Prima facie, if we have this intelligence digitally stored, we should be able to report on it when asked. For example, when questioned: "Could Anne (the person of interest) have been in Edinburgh last Saturday?" We should be able to answer: "Yes, possibly." The use case appears simple, and we might expect that building an information system to hold this information – and provide the right answers would be easy. We show that this simplicity is deceptive by illustrating how the use case raises a series of difficult technical formal challenges that need to be overcome by any information system before it can give a satisfactory answer.

The project was set up to take as its starting point the top-level ontology of the UK Government standard, the Information Exchange Standard (IES) which is based upon the BORO Foundational Ontology [1]). It aims to provide an ontologically precise representation of informational intelligence uncertainty using a two-dimensional approach based upon David Lewis's work on possible worlds and counterpart theory. One designed to be easily implemented on standard IT resources. In the paper, we show how this approach can provide a resolution to all the basic challenges.

## II. Structure of the Paper

We start the paper with an overview of the context for the project. We then set out the challenges facing those (including us) wishing to formalize intelligence uncertainty, centered around our simple use case. We then walk through the overall approach and the technical 'innovations' we are designing to resolve the challenges. We show how the 'innovations' depend upon having a form that can capture and so digitally (formally) express the uncertainty. Together these two sections should help explain and illustrate the formal issues involved. We then take a brief look at the kind of architecture systems that use this work would need to have. Finally, we summarize the paper.

## III. Context – Overview

In this initial section we provide the context for the project. We first frame the search for a formal infrastructure for intelligence uncertainty as providing a tool to reduce a gap in digital maturity. We then give a brief introduction to the project and finally we sketch the foundation upon which the project is building.

### A. Background – Intelligence and Digitalizing Uncertainty

Institutional intelligence can be regarded as the institutional equivalent of personal knowledge – a point made by Kent in *Chapter 1 – Intelligence is Knowledge* in *Strategic intelligence for American world policy* [2] and Heuer in *Psychology of Intelligence Analysis* [3]. Miller [4] highlights three possible senses of institutional intelligence (which also maps to personal knowledge), he says there is "*the threefold distinction between intelligence as the informational, cognitive or epistemic product of intelligence activity, as opposed to the activity itself and the agent (whether an individual or organization) of the activity.*" In this paper, we are interested in '*intelligence as the informational ... product of intelligence activity*', what we will shorten to informational intelligence (though we recognize this is related to activities and agents). One of the key characteristics of this kind of information is the high levels of uncertainty.

Information is uncertain for an agent when they do not know whether it is true or false (see Dubois [5]). This is simple and clear but does not tell us much about the ways uncertain information can indeed be uncertain. Costa et al. [6] see uncertain information as imperfect in some way, including being incomplete, inconclusive, vague or ambiguous. As this list indicates, information can be uncertain in a range of ways that influence how effectively it can be used.

Informational intelligence (and knowledge) has evolved with the human race. Personal knowledge has existed since the dawn of humanity. Institutional intelligence has been around since institutions emerged in the ancient world [7]. More recently, bureaucratization in the late 19th and early 20th centuries, along with other factors including rapid technological advancements and the increasing complexity of economic relations pushed larger institutions (especially states) to not only develop permanent, centralized dedicated intelligence services but also to organize them in a far more systematic way. So, today most industrialized nations' larger institutions have some form of dedicated intelligence service. With these shifts in size and structure institutional intelligence has evolved into something more sophisticated than personal knowledge, though this has evolved too.

Given the nature of their work, it is not surprising that intelligence services generally have a positive attitude towards technology, viewing it as a crucial tool for enhancing their capabilities and addressing evolving security challenges. They typically take a keen interest in new technologies and proactively adopt them. In the case of digital technologies, they recognize that rapid adoption and integration of these is of critical importance in modern intelligence operations and future effectiveness. Where centralized, dedicated intelligence services have significant resources, they are well-placed to act upon this positive attitude to technology. Hence, there are areas where these intelligence services' technology – including digital – capabilities are sophisticated, often extremely sophisticated.

*B. Background – Architectures for dealing with uncertainty*

However, one area that currently represents a challenge is digitalizing the uncertainty itself – finding a form for digital uncertainty. We can distinguish between informational intelligence systems and other more operational systems by their architecture for dealing with this challenge. Many operational enterprise systems (for example, look at SAP, Oracle ERP or Salesforce) pragmatically adopt a policy of excluding uncertainty. Where they (as far as possible) deal with uncertainty before ingesting information into their controlled environment. In this way, uncertainty is mostly banished from their walled garden – and where it occasionally occurs it can usually easily be remedied. In this way, they avoid the need to digitalize it.

Accounting and legal systems provide good examples of this practice. There is an old accounting adage "accountants don't use erasers" that illustrates this. Once a transaction is accepted into the walled garden, it is in a sense certain and stays in the garden. If it needs changing, another reversing entry is added, the original is undisturbed. Legal principles like *res judicata* (Latin – 'a matter judged') where a final judgment is made and cannot be relitigated provide a similar level of certainty.

Informational intelligence systems cannot adopt this strategy as their target subject matter includes the uncertainty itself, so it needs to be ingested and managed. These can take advantage of a range of frameworks and pragmatic tools for formally managing uncertainty. These include probability and decision theory as well as Bayesian networks, fuzzy sets and Monte Carlo networks (see, for example, [6] or Oracle's Primavera P6 system). Rather than attempting the challenging work of representing uncertain information directly, such systems focus on pragmatic workarounds to represent aspects of information of interest. While there is of course pragmatic value in leveraging such strategies when dealing with uncertain information, there is also value in wrestling with more direct representations of uncertainty, such as what one might find in the context of an *ontology*. When ontologies include data, moreover, they often exhibit rich semantics, insofar as they make explicit formal structures that are often left implicit in and across datasets. The focus of this project is the development of an ontology-based digital form (aka formal data infrastructure) for uncertainty. The expectation is that then some forms of uncertainty will become significantly more tractable, owing to the clarity, precision, and richness of our ontological representations. For the users, this tractability will translate into with more expressivity.

*C. Background – Project to Digitalize Uncertainty*

Our project aims to digitally formalize informational intelligence uncertainty. It takes as its starting point, its foundation, the top-level ontology of the Information Exchange Standard (IES), the BORO Foundational Ontology [1] – described in the next section. The aim of this project is to design and test a formal digital infrastructure extension to the BORO/IES foundation which will provide an ontologically precise representation of informational intelligence uncertainty, with the aim of promoting stakeholder use and understanding of informational intelligence uncertainty. This will be incorporated into the IES standard as well as the BORO Foundational Ontology ecosystem. In this way, we hope to demonstrate that it is feasible to digitally formalize this uncertainty and show what the formalization looks like. A future project will address the challenge of socializing this new capability.

In the project we adopted an empirical Information Systems (IS) flavored approach to the development of the formal digital infrastructure. We started by defining the scope of uncertainty in terms of use cases. We are using BORO's tools (and experience of) mining ontological commitment from data to develop and test resolutions to these use cases. We are also using relevant philosophical research to guide our analysis. Currently, we are co-evolving use cases and resolutions to both deepen and clarify our understanding of issues and enhance the sophistication and resilience of the infrastructure.

*D. Background – BORO and IES Ontology*

We take as the starting point or foundation of the project the top-level ontology of the IES which is based upon the BORO Foundational Ontology. This has been found to be extremely useful in many ways, including its clear ontological identity criteria (as explained in [8]).

IES is a standard for information exchange developed within the UK Government. It is based upon the BORO Foundational Ontology (often just shortened to BORO, an acronym for 'Business Object Reference Ontology'). BORO is one of the earliest top-level information system ontologies. Its development and deployment which started in the late 1980s as is described in *Business Objects* [9]. BORO's focus was and is on enterprise modelling; more specifically, it aims to provide the tools to salvage the semantics from a range of enterprise systems building a common foundation in a consistent and coherent manner. The work we present in this paper is built upon a BORO/IES foundation which is briefly sketched below – but well-documented elsewhere (see for example [1]).

BORO is grounded in philosophy and has clear meta-ontological choices [10] following paths well-established in twentieth and twenty-first-century philosophy [11], particularly those found in the philosopher David Lewis's mature work, especially *On the Plurality of Worlds* [12]. This makes BORO a good foundation upon which to build Lewis's approach to uncertainty. BORO's choices are categorized and compared with other top-level ontologies in *A Survey of Top-Level Ontologies* [8]. Over the last decade, BORO has been enhanced with a constructional approach that clearly reveals its parsimonious foundations. One where the whole ontology can be constructed using three constructors (set, part and tuple), starting with a single object – the pluriverse [1].

BORO adopts extensionalism, so at the level of mereology it accepts not just that everyday things are extended in both space and time – and can be visualized as four-dimensional worms coexisting in spacetime – but that this extension is the basis for identity. The person Anne and the city Edinburgh are both four-dimensional worms extended in spacetime, as are other people and cities. Times are also four-dimensional worms – 30th March 2024 is a timeslice of the whole universe for the relevant period. This means that spatial and temporal locations end up as simple mereological relations between worms. The event of Anne being in Edinburgh would be a temporal slice of Anne (the four-dimensional worm) that overlapped (a mereological relation) with Edinburgh (the four-dimensional worm) giving a smaller four-dimensional worm that was part of the bigger Anne and Edinburgh worms. If Anne was in Edinburgh on 30th March 2024, then the Anne in Edinburgh worm would be a part (a mereological relation) of the 30th March 2024 worm. This simplifies a lot of things into mereological structure.

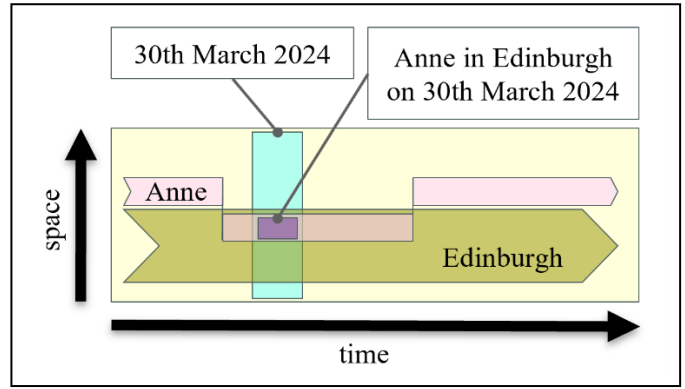

Fig. 1. Space-time diagram: Anne in Edinburgh on 30th March 2024

BORO also adopts 'possible worlds', more specifically the existence of every kind (the plenitude) of possibility, so a more precise name for the choice would be 'possibilia plenitude'. In the design-space of top-level ontologies, the possibilia plenitude choice enables property extensionalism allowing any property to be identified extensionally with the set of all possible individuals with that property (explained in, for example, [13] as well as [8]). In other words, the property set contains (that is, have as its extension) all the ways in which the property could possibly exist. The extension of the set provides the identity criteria of the property. In other words, a given property is defined entirely in terms of its extension.

Modality (sometimes called alethic modality when it needs to be distinguished from other modalities) is traditionally about possibility: where the topic is divided into the properties of possibility versus impossibility and within possibility, into necessity versus contingency. Possible worlds open up two new ways of looking at modality, which has been characterized in terms of dimensions [14], [15], [16] – also Lewis's [12]. Firstly, a one-dimensional approach that allows for simple expressions of modality and then a two-dimensional approach, *de se* indexed (or centered) on a context that allows for more sophisticated expressions of modality.

Another way of thinking about this is that the modalities can be divided into those about properties ("Necessarily, all dogs are mammals") and those about individuals ("Necessarily, Fido is a dog"). These are sometimes distinguished respectively [13] as *de dicto* and *de re* (Latin for 'about what is said' and 'about the thing', also respectively) modalities, though one needs to be careful as these two terms — have a bewildering range of related senses.

Out of the box, the 'possibilia plenitude' and 'extensionalism' choices enable one dimensional (so-called) *de dicto* property modalities to be explained structurally in terms of the extensions of the property sets – the modalities emerge from the structure of the sets. 'Possibly, some dogs are mammals' says the property set dogs overlaps with the property set mammals. 'Necessarily, all dogs are mammals' means the property set dogs is a subset of the property set mammals. 'It is impossible for a cat to be a dog' means the property set cats does not overlap with the property set dogs. 'Contingently, some dogs are male' means the property set dogs overlaps the property set males – but is not a subset of it. The modality arises from the extensional relations between the property sets.

*De re* individual modalities need more structure – the ability to link individuals across worlds. Tackling this is one of the challenges of this paper. We shall see later how the adopted approach, like the property modality approach, reveals individual modalities with a mereological and set-theoretic structure.

## IV. BASIC CHALLENGES

### A. Should handle de re individual modality

As noted earlier, intelligence uncertainty often involves *de re* individual modality – where this focuses on a specific individual. This is different from the other form of modality we noted earlier, de dicto property modality – this kind of claim would be that: a person can be in a city at a time. Our use case exemplifies the *de re* individual kind of modality, it is about the specific individual Anne – the person of interest. The challenge is to find a way of expressing the modal properties of this individual; expressing that it is possible that Anne could have (*possibly*) been in Edinburgh on 30$^{th}$ March 2024.

### B. Should handle actuality

Our talk is often implicitly about actuality. When we say Anne might have been in Glasgow yesterday, we usually mean she might *actually* have been there. Not that if she had arranged things differently, she could have been there. Informational intelligence is, at its core, usually the same – it involves information about what has, is or will, may or might, *actually* happen. One can contrast this with counterfactual simulations that evaluate how different decisions or events could have led to different outcomes, where there can be little interest in whether the events are actual, rather a positive intention that they are simulating an alternative to what *actually* happened. If one develops a form for expressing possibility, one needs to ensure that within the notion of possibility there is a way to express actuality.

Our use case exemplifies both this actuality and its implicit assumption. From the last challenge, we can say that Anne might *possibly* have been in Edinburgh on 30$^{th}$ March 2024. But now we want to know (and say) whether Anne might *actually* have been in Edinburgh on 30$^{th}$ March 2024. We want to be able to say that it is possible that she was *actually* there. And also, to be able to exclude cases where we want to say it was merely possible for Anne to be in Edinburgh if she had, for example, rearranged her diary, but as she did not, she was definitely *actually* not there. The challenge is to find a formal way of expressing this actual possibility.

### C. Should handle 'inconsistent' knowledge

Imperfect, even (apparently) inconsistent, knowledge is a common human predicament and often a feature of intelligence. For example, we may have some information that says Anne might have been in Edinburgh and other information that she might have been in Brighton. It is obvious she could not have been in both places at the same time. We think both bits of information are plausible, but only one can be correct. So, then it is possible, but not definite, that Anne was in either place but not in both. What form do we use to express this?

### D. Should be owned

There is not just one intelligence, there are many different intelligences, and each is someone's intelligence. As [4] says: "intelligence is … institutionally relative, (i.e., relative to some institutions)" A recent example is President Biden's 90-day covid pandemic origins review [17] which said: "Four IC elements and the National Intelligence Council assess with low confidence that the initial SARS-CoV-2 infection was most likely caused by natural exposure to an animal infected with it or a close progenitor virus ... One IC element assesses with moderate confidence that the first human infection with SARS-CoV-2 most likely was the result of a laboratory-associated incident ... Analysts at three IC elements remain unable to coalesce around either explanation without additional information ..." So, each element had its own intelligence. This example also illustrates several of the other challenges, including the limited information introduced above.

The President clearly knew which elements the answer came from. So, for the use case, we should be clear which intelligence element is involved – and when we are asked, we should get an answer which contains a marker identifying the provenance. There is also a stronger requirement for us to be able to say which is our intelligence (what is sometimes called *de se* (Latin – 'about oneself')). Otherwise, when we get further intelligence, we will not know which intelligence to update. So, a requirement is for our intelligence system to know that our intelligence is that: Anne could have been in Edinburgh on 30$^{th}$ March 2024 and at the same time know that another agent's intelligence is that: Anne could have been in Brighton on 30$^{th}$ March 2024.

### E. Should handle currency

If informational intelligence is current, then it changes over time. One well-known example is the Central Intelligence Agency's (CIA) changing assessment of the Soviet Union's economy and military capabilities during the Cold War. In the 1950s and early 1960s, the CIA generally overestimated Soviet economic and military strength. By the 1970s and 1980s, the CIA's view began to shift, recognizing significant weaknesses in the Soviet economy and military. This shift in perspective contributed to changes in US policy and strategy towards the Soviet Union, ultimately influencing the approach that led to the end of the Cold War. This illustrates that intelligence assessments are made at a point in time, current for that time. That what is current changes over time and that these changes can have significant impact.

In terms of our use case, the information store needs to be expressive enough to identify current intelligence – the information in our store that is valid now ('now' is sometimes called *de nunc* (Latin 'about now')). This implies that the intelligence store needs to have a form that can express now (*de nunc*). If it stores information about Anne being in Edinburgh on 30$^{th}$ March 2024, it needs to be able to mark this as current. If it no longer thinks the information is valid, so it is no longer current, it needs to be able to have a 'memory' of what it used to 'think' was current. We return to this in a later challenge.

*F. Should be modally consistent*

As noted in a previous challenge, good intelligence can be inconsistent, in the classical sense, where two pieces of intelligence cannot both *actually* be true. But it should be modally consistent. For an example, we return to President Biden's 90-day covid pandemic origins review. This noted that some of the services "*do not believe there is sufficient information to assess one to be more likely than the other.*" Prima facie, they assume that either the zoonotic or the laboratory leak scenarios could possibly be true. This is inconsistent in the classical sense, as both scenarios cannot both be true *simpliciter*. However, it is not inconsistent in the modal sense, as (explaining using a 'possible worlds' stance) both scenarios could be true in their own possible world. Obviously, they could not be true in the same possible world, as this would be modally inconsistent.

In terms of our earlier use case of Anne being in Edinburgh or Brighton. The information store needs to have an expressive framework that enables it to store the two pieces of information in a way that is modally consistent and correctly answer questions based upon this. And it needs to have guard rails, boundaries, that would highlight cases that are modally inconsistent – such as information that Anne is in Edinburgh and Brighton at the same time (in the same possible world).

*G. Should handle credence relations*

Informational Intelligence has a network of credence relations. In this network, there are simple objective dependency relations between pieces of informational intelligence – where one in a sense 'contains' the other. If the contained information is true, then the containing information is necessarily true. This dependency can be captured with a strict conditional – in Lewis' [18] sense. These are objective in the sense that the dependency is not relative to a particular system. If two intelligence systems agreed on the information, they would agree on the dependency. For example, in the 2003 Iraq War (Operation Iraqi Freedom), U.S. intelligence gave high credence to the general suggestion that Iraq possessed Weapons of Mass Destruction (WMD) and was seeking to develop more. One specific piece of intelligence, which came from various sources including an Iraqi defector codenamed "Curveball," claimed that Iraq had mobile biological weapons labs [19]. Given that mobile biological weapons labs are weapons of mass destruction, if the more specific claim is true, then it follows that the general one is too, but not vice versa. These kinds of dependence need to be at least represented and, if possible, explained by the formal information infrastructure.

If there is necessity in one direction (from the contained to containing) there is relative possibility – and so credence – in the other direction. It is possible that there were no mobile biological weapons labs, but Iraq did have WMD. In this situation agents will naturally assess the credence of the more specific (mobile labs) relative to the credence of the more general (WMD) – creating a dependency (see Ramsey's [20]). For example, the U.S. intelligence assessment initially gave a high credence to the specific claim of mobile biological weapons labs despite its lower verifiability probably based upon the high credence given to the general claim. It subsequently reduced its level of credence when it also lowered its credence for the general claim. In a simple way this illustrates how this dependence plays a role in the network of credence relations. Typically, this assessment is subjective, in that different agents can give different assessments.

For our use case, we could have a general assertion that 'Anne was in Edinburgh on 30th March 2024' and the more specific assertion that 'Anne met Effie in Edinburgh on the 30th March 2024'. There is a clear objective dependency between the specific and general assertions – if Anne met Effie in Edinburgh, then she necessarily was in Edinburgh. So, it would be odd to claim the specific assertion was true and the general assertion false. We would also expect their credences to be related. If we change our credence that Anne was in Edinburgh from high to low, we expect some corresponding reduction in our dependent credence that Anne met Effie in Edinburgh. We have expressed this as a ranking of credences. One could associate probabilities rather than rankings with these credences.

*H. Should handle testimony*

Informational intelligence is based upon a network of testimony, memory and assertions. As noted earlier, analogies with everyday knowledge can be insightful. In everyday life, when we make assertions [21], [22] we would like them to be based upon easy to justify first-hand experience, especially immediate experience. However, we find that we often need to rely on the less easy to justify testimony (assertions) of others.

Our memories, it has been argued, are a kind of testimony from our past [23], [24]. And similarly argued that in some ways our reliance on memory is like our reliance on testimony. Whatever the similarities and differences, we find that we rely on our memories of first-hand experience and of the testimony (assertions) of others. These others are in a similar position, having themselves to rely on memory and testimony and so networks of memory and testimony are created. Plainly, these networks are both an indispensable source of knowledge as well as useful in assessing the credence we should give to that knowledge.

Not all testimony is equal. We need to decide whether to accept, how much to believe, the testimony we receive. And how we judge is influenced by the context in which we receive the testimony, who gives it and how they give it. By analogy, we can see a similar network is a core feature of intelligence operations. This creates a challenge for an informational intelligence system. It needs to be able to store testimony as testimony down a testimony chain as well as the information testified and manage the judgements of its validity.

There are a range of other interesting analogies, we just highlight one: memory reconsolidation [25], where memory retrieval leads to changes in the memory trace. We thought Anne might have been in Edinburgh, when we get reliable information that she was in Brighton, we not only need to record that but also update our memory that Anne might have been in Edinburgh. This is a requirement – and so a challenge – for informational intelligence systems.

*I. Should handle flexible system identity*

Digital informational intelligence systems should have more flexible identity. Digitalization typically brings a range of general benefits, such as improvements in speed, accuracy and quality as well as scalability and the potential for enhanced security while also reducing costs. There are also potential benefits specific to informational intelligence systems, one of which we focus on here – the facilitating of branching (fission) and merging (fusion) of informational intelligence. This will be required when, for example, intelligence organizations merge or split, and their systems need to too. To see the issue, it helps to consider the human analogue from a general design perspective rather than the physiological details. The philosophy of personal identity looks at the conceptual challenges associated with whether persons can undergo fission and fusion [26], [27] raising interesting questions about bodily and psychological continuity. The informational intelligence systems face a challenge in meeting similar, but less stringent, requirements for bodily (physical) and informational continuity in their network of testimony and memory.

V. RESOLUTION

In this section, we outline our current work on the resolution of these challenges. We first provide a sketch of the overall approach and then work through the challenges.

*A. Overall Approach*

As noted earlier, the BORO/IES foundation is based upon David Lewis's mature work and so has basic foundational elements already in place, including the metaphysical choices of extensionalism and possibilia plenitude. Taking Lewis as our guide, we extend these to meet the challenges of designing a form for uncertainty in informational intelligence.

*1) Architectural choices – use standard resources*

Lewis [28], [29] quite explicitly from the start says that he adopts a strategy of formalizing modality using standard resources (such as first order logic) without the use of specialized modal operators (such as the boxes and diamonds of modal logic). He claims this more direct approach has many benefits including being both more explanatory and expressive.

By adopting Lewis, we inherit his strategy. This gives us what might be called from a computing perspective a pure object-oriented perspective. Object oriented because simple modal statements about possible individuals or properties are translated into statements directly about objects. So, the statement 'the individual $x$ is possible' becomes directly '$x$ exists', where '$x$ is part of some (possible) world'. Then when we want to assess a possibility, all we need to look for is the 'possible' object. Pure in that, unlike object-oriented programming objects, there is no paraphernalia of attributes and methods. Lewis was clear on the benefits of sticking with standard resources [28], and these benefits carry over to the digital implementation, where questions of possibility can then be answered with suitable forms with standard enterprise computing resources such as relational databases or OO programming languages. This simplifies technological issues of performance, scaling and reuse.

*2) Architectural choices – two-dimensional architecture*

Another feature of Lewis that we adopt is his two-dimensional architecture (see earlier references on this kind of architecture) for his two key context indexicals – where indexicals are signs that point indexically, in the sense that what they point to depends upon the context of utterance. Classical indexical linguistic pronouns are 'I' which indexically refers to whomever is speaking and 'now' which indexically refers to the moment at which it is spoken. In philosophy, these correspond with respectively *de se* corresponding to 'me' or 'I' and *de nunc* corresponding to 'now'. Lewis characteristically builds these context indexicals into the model.

There is a difference in approach due to the nature of our project. In the academic literature there is an understandable focus on the visible public utterance of a speaker – as the contents of the speaker's mind are invisible, private. There is an asymmetry with our situation, where we are designing the form of the information store, so this is not just public, but the design is under our control. In the academic literature, the context is a speaker making an utterance at some point in time, which Lewis characterizes as *de se* and *nunc*. The focus is not on the inside of the speaker's mind which, presumably, privately contains the content, the representation of the world (ontology). In our informational intelligence case, the focus is rather directly on the current centralized, controlled systems that store information and only indirectly on the system's response to a query, which would have a similar role to the utterance. There is a broad similarity, in that the system plays the same role as the speaker, and the information store plays the role of content, which is used to represent the world (ontology). Our project is to design a suitable form for this store. As you will see in the resolutions below, we do this in large part by adapting Lewis's architecture to this different kind of context.

*3) Use case testing*

We plan to implement the use cases in a test informational intelligence system and so demonstrate the challenges being met.

*B. Facing the Challenges*

We take the challenges from the earlier section in turn, explaining how we design a form that resolves them.

*1) Introducing individual properties*

The *de re* individual modality challenge is to find a form for individual possibility (described earlier and contrasted with property possibility). To, for example, find a form for saying it is contingent (possible but not necessary) that Anne was in Edinburgh on 30th March 2024 that allows it to be possible that she was in Edinburgh, and also that she was not.

*a) Different possible Annes*

Given the possible worlds approach, this means that there exists both a possible Anne who was in Edinburgh and a possible Anne who was not. These need to be different as it is impossible to both be and not be in Edinburgh at the same time. And given we believe there is no more than one Anne in this (or any other) world, then these different Annes must be in different worlds. In our foundation, these are different, unconnected objects – and there is not yet the machinery to connect them. So, the task is to find a way of connecting these (trans-world) Annes.

*b) Individual properties*

Lewis's answer is to connect the objects with a counterpart relation [28]. He stresses the flexible way in which counterparts work, suggesting people select the counterparts they need for a given situation. This counterpart relation picks out a set of related individuals – which we call the individual possibility.

We shorten this to possibility where we have an individual prefix, for example, 'Annes individual possibility' becomes 'Annes possibility'. We may shorten this even further, where the prefix is sufficiently informative, so for example 'Annes possibility' becomes 'Annes'. As sets (of individuals), individual possibilities are Lewisian properties – the individual possibility properties. (Lewis talks about similar event properties [30], though for a different purpose.) After looking at a variety of different approaches, we have found, so far, that for our informational intelligence purposes it makes sense to work with individual possibilities without identifying the individual members. So, for example, we would recognize 'Annes possibility' as a property – a set – of all the possible (individual person) Annes. We have similar properties for the Edinburgh possibilities and the 30th March 2024 possibilities. This assumes we can devise stable individual possibilities for the kinds of systems we want. Whether we can is an empirical matter and needs to be tested. The first test being our use cases.

*c) Individual properties' properties*

Once we have these individual possibilities, we can build a system of modal properties on them. A set of individual possibilities will have the property of being either compossible (jointly possible in some world) or incompossible (jointly impossible in all worlds). If compossible they will have the property of being either comnecessary (always jointly possible) or comcontingent (only sometimes jointly possible). In our use case, there is a further modal property and associated construction that is useful: this is comoverlapability (somewhere jointly overlap), where compossible individual possibilities overlap. Consider the set containing the two individual possibilities Annes and Edinburghs. Let's say it is compossible, so there are some worlds that contain both an individual Anne and an individual Edinburgh. The individual possibility is comoverlapable if any of the Annes and Edinburghs overlap – in other words, if Anne ever visits Edinburgh. Compossible properties can also be used to construct their associated individual possibilities using mereology. In the case of comoverlapability, we construct the new individual overlap possibility from the individual possibilities that are overlapping – in this case, the states of Anne being in Edinburgh. We can add the 30th March 2024s possibility to the mix and comoverlapably construct the Anne being in Edinburgh on 30th March 2024 possibility. We know this individual possibility must exist as we have a 'possibilia plenitude' – anything that is possible exists. As the example shows, comoverlapability gives us the tools to talk about an individual possibility's spatial and temporal locations.

*d) Resolving de re individual modality*

Individual properties give us the means to resolve *de re* individual modality. A way of expressing that it is possible that Anne could have (possibly) been in Edinburgh on 30th March 2024 – but might not have been. If we take the Annes individual property, some of its members are in Edinburgh on 30th March 2024, others are not. This captures *de re* individual modality.

2) *Resolving actuality*

In the Lewisian possible worlds, the actual world is *de se* indexed – it is the world of which I am part. In other possible worlds, there will be people for whom their world is the actual world. In Lewisian indexicality, the actual world is picked out by the human speaker making the utterance – hence *de se*. It is made explicit when the speaker uses the indexical pronoun 'I' in the utterance – referring to herself. Parts of the speaker's world are actual. All other objects are not. In our informational intelligence case, we have an information system at the center rather than a speaker. We can make the indexical explicit by adding a 'me' sign to the system that refers to itself, something some of the authors have discussed elsewhere [31], [32]. With this infrastructure, one can represent formally an information system recording 'Anne might actually have been in Edinburgh on 30th March 2024' by recording that the 'Anne was in Edinburgh on 30th March 2024 possibility' and the singleton actually me possibility are compossible – which is another way of saying the set of these two is a member of the compossibility property. This resolves the informational intelligence as actuality challenge.

*a) Resolving ownership*

If all information is *de se*, then it belongs to someone – the 'I' who has the information. By explicitly introducing the sign for 'me' into the system, we clearly expose the 'owner' of the information. This resolves the 'intelligence belongs to someone' challenge – it belongs to the information system.

3) *Introducing doxastic actuality*

Resolving the next few challenges requires designing a Lewisian doxastic structure for informational intelligence which involves a series of steps.

The first step is to introduce the capability to represent *de nunc* actuality. For this we add to the information system an indexical sign for 'my actuality' – which refers to me (the information system) *now* – the timeslice of me now (BORO had an early version of this [9]). The second step is to introduce what Lewis calls 'doxastic alternatives':

*"We should characterise the content not by a class of possible worlds, but by a class of possible individuals – call them the believer's doxastic alternatives – who might, for all he believes, be himself. Individual X is one of them iff nothing that the believer believes, either explicitly or implicitly, rules out the hypothesis that he himself is X. These individuals are the believer's doxastic possibilities."* [12, pp. 28–9]

We call these my doxastic actualities. It takes a few steps to formalize this. We firstly look at the individual possibility of the *de se* and *nunc* 'my actuality' – constructed comoverlapably (as described earlier) from the individual possibility of me with the (*de nunc* indexed) now individual possibility. This set contains as members all the individuals that could possibly be my actuality - me now. We want to filter these to 'my doxastic actualities': those of my actuality individual possibility members who hold the same beliefs as me and whose beliefs are compatible with themselves.

We have a choice here. We could reify individual possibilities relative to the my doxastic actualities possibility. We consider the my doxastic actuality worlds possibility, those worlds that contain a member of the my doxastic actualities possibility, We then restrict possibilities to these worlds. The Annes possibility is then restricted to the doxastically actual Annes possibility. These individual possibilities can be volatile.

Instead, we capture the compatibility restriction by setting up in the system the individual doxastic actuality properties shown in TABLE I. which also notes the relation their members have with 'my doxastic actualities' properties.

TABLE I.   INDIVIDUAL DOXASTIC ACTUALITY PROPERTIES' RELATIONS

| Property | Relation |
|---|---|
| possible | compossible |
| necessary | comnecessary |
| contingent | comcontingent |
| impossible | incompossible |

When the system has a belief, for example that it is contingent Anne was actually in Edinburgh on 30th March 2024, we need to capture that the 'Anne was in Edinburgh on 30th March 2024 possibility' is compossible with 'my doxastic actualities' – in other words, a member of the relevant individual possibility property.

#### a) Resolving 'inconsistent' knowledge

Let's say we decide the system also believes that it is contingent Anne was actually in Brighton on 30th March 2024. Again, we need to note in the store that the 'Anne was in Brighton on 30th March 2024 possibility' is compossible with 'my doxastic actualities' – in other words, a member of the relevant property. This resolves the 'intelligence as inconsistent knowledge' challenges.

#### b) Resolving currency

Over time beliefs of what is actual change. This is reflected in the information system by changing the 'my doxastic actualities' compatibility restrictions. We show the system now believes that it was impossible that Anne was actually in Edinburgh on 30th March 2024 by changing the membership of the 'Anne was in Edinburgh on 30th March 2024 possibility' from compossible to incompossible. This resolves the Informational intelligence is current challenge.

#### c) Resolving modal consistency

The system has a 'my doxastic actualities' possibility property with (typically) several members. This creates space for inconsistent possibilities. If we want to store the information that Anne could have actually possibly been in either Edinburgh or Brighton, this is cashed out as she is in Edinburgh is some of the system's my doxastic actualities possibility member worlds and in Brighton in others. It is true in some of my doxastically actual worlds that she is in Edinburgh and also true in some of my doxastically actual worlds that she is in Brighton and that in none of these worlds she is in both (at the same time). This resolves the 'informational intelligence should be modally consistent' challenge.

#### 4) Resolving credence relations

In our use case, we make a general assertion that Anne was in Edinburgh on 30th March 2024 and the more specific assertion that: Anne met Effie in Edinburgh on 30th March 2024. In our resolution, these correspond to two possibilities: the 'Anne was in Edinburgh on 30th March 2024 possibility' and the 'Anne met Effie in Edinburgh on 30th March 2024 possibility'. These have a clear structural relationship. Some of the members of the first overlap with all of the members of the second. So, in cases where the second is possible, the first is also possible. In this approach, modal dependency becomes mereological dependency. More generally, often modal dependency becomes structural mereological and set-theoretic dependency. This is objective in the sense that the relation is not dependent in any way upon my doxastic actualities.

These dependency relations give an order over the credences – enabling us to say one credence is more or less likely than another. Lewis has done much work showing how a Bayesian epistemology of changing credences would fit into his architecture, which we are working on at the moment. One can see the beginnings of this in the simple case of Anne and Effie above. The relative credence would be between the two Anne and Effie possibilities – in the context of my doxastic actuality possibility. A particularly interesting suggestion is that the relative credence is attempting to measure the ratio between the my doxastic actuality members of the Anne in Edinburgh possibility and the Anne met Effie in Edinburgh possibility. This resolves the network of credence relations challenge.

#### 5) Resolving testimony

Once we have the basic centered pattern, we can reuse it for all the doxastic structures. To store information about a network of testimonies, we just need to recreate the centered pattern at each node. This is a little convoluted, but can be illustrated with an example. Consider a case where our information system records the testimony that Bindi said that Anne could have been in Edinburgh. We first need to recognize Bindi. We construct the Bindis possibility – unindexed to any doxastic alternatives. And then index it as actually comnecessary to the information system – assuming it doesn't doubt her existence. Then the system's (my) doxastic actualities possibility will always share worlds with worlds a member of the Bindis possibility. We look at the Anne in Edinburgh possibility – again unindexed to any doxastic alternatives. We then consider the Bindis possibility's doxastic actualities at the time of testimony – which will include the belief that the Anne in Edinburgh possibility is necessary.

This structure captures that the system accepts that Bindi exists and that she believed at the time of the testimony that Anne was in Edinburgh but is neutral about whether the system believes this as well. This indicates how to resolve the network of testimony, memory and assertions challenge.

#### 6) Resolving flexible system identity

As we have a reasonably clear picture of the semantic form of the information, then the splitting and merging of the informational intelligence system should not be an insurmountable challenge. Once we have implemented the system using the use case, we should be able to test our ability to do this. Lewis [26], noted earlier, suggests that we will need to be precise with the semantics of 'me'. In the case of branching and merging, there may be one 'me' stage that is part of two different 'me's. This should not be an insurmountable problem. When we run our use case tests, we should be able to show how branching and merging is handled.

## VI.  CONTROLLED INFORMATION SYSTEMS

As noted earlier, informational intelligence systems opt for an architecture where the uncertainty is ingested and managed inside the 'walled garden'. This walled garden is a controlled system in the sense of having a boundary within which there are patterns or rules of behavior that are followed – to maintain a holistic structure. The formal structure of automation enables internal rules that barring accidents the system will always follow – so controlled systems. The overall ecosystem includes human users who will need to respect and follow these rules.

The resolutions above use an ontology-based digital form (aka formal data infrastructure) to represent uncertainty, unlike that used in current systems. As the infrastructure is formal, there need to be controls to ensure consistency. So, we will need to design how an informational intelligence system using this infrastructure would work. These may be similar to existing controls in the more manual systems but are unlikely to be so across the board. Hence, this is likely to need some trial-and-error testing – and we should recognize this from the outset.

## VII. Conclusion

The paper starts by describing the challenges dealing with uncertainty faces. It then describes how this project is facing them. It shows how an extensional ontology (such as BORO/IES) can be extended with a Lewisian counterpart approach to formalizing uncertainty in a way that is both adapted to computing and expressive enough to handle the challenges.

As Lewis has noted, the uncertainties of knowledge need a flexible approach, one that sometimes even seems a bit sloppy. But as Lewis has also noted, and as we demonstrate here, this does not mean it cannot be given a clear formal framework. The next stage of the project is to demonstrate, using the use cases, how the framework makes managing some aspects of uncertainty more tractable.

The framework is unabashedly extensional – cashing out as a combination of set-theoretic and mereological relations – which gives it a comforting explanatory feel. For example, the actual world is the world that I am part of (mereology). And the modal property 'compossible' means that members (set theory) of the counterparts are jointly part (mereology) of some world. Where being possibly spatially or temporally located is just the modal property of comoverlapability – where this cashes out in the same extensional way. The simplicity that emerges from explaining uncertainty through this extensional lens feeds into the framework. And the project aims to show how this leads to more tractable treatments of uncertainty.

## Acknowledgment

We wish to acknowledge the helpful funding of the project by the UK Government and the extremely useful input by all the collaborators on this project. We also thank the reviewers A. Cooke, S. de Cesare, J. Beverley and M. Farrington.